\documentclass{ws-procs961x669}            

\newcommand\araa{{ARA\&A\,}}		
\newcommand\apj{{ApJ\,}}			
\newcommand\aap{{A\&A\,}}			
\newcommand\mnras{{MNRAS\,}}		
\newcommand\pasa{{PASA\,}}			
\newcommand\nat{{Nature\,}}			

\newcommand\prd{{Phys.~Rev.~D\,}}

\newcommand\ssr{{SSR}}

\newcommand{\dif}{{\rm d}}
\newcommand{\Ms}{\ensuremath{\mathrm{M}_\odot}}
\newcommand{\Rs}{\ensuremath{\mathrm{R}_\odot}}
\newcommand{\Mpy}{\Ms\,{\rm yr}{\ensuremath{^{-1}}}}

\begin{document}

\title{The Rotation of SuperMassive Stars}
\author{L. Haemmerl\'e}
\address{D\'epartement d'Astronomie, Universit\'e de Gen\`eve, chemin des Maillettes 51, CH-1290 Versoix, Switzerland\\
E-mail: lionel.haemmerle@unige.ch}

\begin{abstract}
Supermassive stars (SMSs), with masses $>10^5$ \Ms,
have been proposed as the possible progenitors of the most extreme supermassive black holes observed at redshifts $z>6-7$.
In this scenario ('direct collapse'), a SMS accretes at rates $>0.1$ \Mpy\ until it collapses to a black hole via the general-relativistic (GR) instability.
Rotation plays a crucial role in the formation of such supermassive black hole seeds.
The centrifugal barrier appears as particularly strong in this extreme case of star formation.
Moreover, rotation impacts sensitively the stability of SMSs against GR, as well as the subsequent collapse.
In particular, it might allow for gravitational wave emission and ultra-long gamma-ray bursts at black hole formation,
which represents currently the main observational signatures proposed in the literature for the existence of such objects.
Here, I present the latest models of SMSs accounting for accretion and rotation, and discuss some of the open questions and future prospects in this research line.
\end{abstract}


\bodymatter

\section{Introduction}
\label{sec-intro}

Supermassive stars (SMSs), with masses $\gtrsim10^5$ \Ms,
were originally invoked to explain the high luminosity of quasars\cite{hoyle1963a,hoyle1963b},
later understood as a consequence of accretion in the deep potential well of supermassive black holes (SMBHs)\cite{lyndenbell1969}.
The formation of these SMBHs remains, however, one of the main open problems in galaxy formation,
and SMSs have been proposed as their possible progenitors\cite{rees1984,volonteri2010}.
In the last decade, a number of quasars powered by SMBHs with masses $10^9-10^{10}$~\Ms\ have been detected at redshifts 6 -- 7.
The most extreme case known to date is J0313-1806 a quasar at $z=7.6$ that hosts a SMBH with mass $1-2\times10^9$ \Ms\cite{wang2021}.
At redshifts 6 -- 7, the age of the Universe is a fraction of a Gyr, so that objects of $10^9-10^{10}$ \Ms\ that exist at these redshifts
must have accumulated their mass at average rates $\dot M\gtrsim10$ \Mpy.
The direct formation of SMBHs during the process of galaxy formation ('direct collapse' black holes)
provides the most rapid pathway to form SMBHs with masses $\gtrsim10^9$ \Ms\cite{volonteri2010,wang2021}.
In this scenario, the direct progenitor of the SMBH is a SMS undergoing accretion at rates $\dot M\gtrsim0.1$ \Mpy,
until it collapses due to the general-relativistic (GR) instability \cite{woods2019,haemmerle2020a}.

The direct collapse of $\sim10^9$ \Ms\ into a compact object requires special conditions.
Fragmentation is inhibited in primordial gas due to the absence of heavy elements,
but the cooling by molecular hydrogen remains efficient enough to maintain temperatures 100 -- 1000 K,
which is thought to result in the formation of a cluster of stars with individual masses $M\lesssim100$ \Ms\cite{klessen2019}.
Black hole seeds formed from the collapse of such primordial clusters remain $\lesssim1000$ \Ms,
and hardly explain the large masses of the earliest quasars\cite{lupi2014,wang2021}.
Only in the presence of a strong Lyman-Werner flux the molecular hydrogen of primordial gas can be dissociated,
which leads to temperatures $\sim10^4$ K set by atomic cooling\cite{omukai2001}.
Primordial, atomically cooled halos are found to collapse isothermally with inflows 0.01 -- 1 \Mpy\ sustained down to sub-parsec scales,
allowing for the direct formation of a supermassive compact object in a dynamical time\cite{latif2013,chon2018}.
However, this scenario implies tight synchronisation between halo pairs,
since only star formation in a neighbouring halo can provide the required Lyman-Werner flux\cite{regan2017}.
Another channel of direct collapse is provided by the merger of the most massive galaxies expected to have formed at redshifts 8 -- 10,
with masses $\sim10^{12}$ \Ms\cite{mayer2019}.
The merger is found to trigger inflows of $\sim10^4-10^5$ \Mpy\ down to a fraction of a parsec,
which allows to accumulate several $10^9$ \Ms\ in a few tens of kyr.
No special chemical composition is required, since the collapse is too rapid for star formation and radiative feedback to start,
and metallicities up to solar are expected in these galaxies.
However, SMBH formation by direct collapse in metal-rich gas requires a SMS with mass $M\gtrsim10^6$~\Ms\
in order to avoid thermonuclear explosion\cite{montero2012}.

The strong contraction that characterises a gravitationnal collapse leads in general to rapidly rotating structures,
by a simple argument of angular momentum conservation.
In fact, star formation by accretion requires strong mechanisms to remove angular momentum from the collapsing gas
in order to circumvent the centrifugal barrier\cite{bodenheimer1995}.
This angular momentum problem represents one of the main bottleneck in models of massive star formation\cite{lee2016,haemmerle2017,takahashi2017}
and plays naturally a critical role in the context of SMS and SMBH formation\cite{begelman2006,haemmerle2018a}.
Here, I review the latest theoretical results concerning SMSs in the context of SMBH formation, focusing on the importance of rotation.
In Section~\ref{sec-sms}, I review the main properties of SMSs formed by accretion;
I discuss the angular momentum problem in section~\ref{sec-J};
section~\ref{sec-gr} focuses on the GR instability and the maximum mass of rotating SMSs;
the implication for SMBH formation by direct collapse are discussed in section~\ref{sec-smbh}.
A summary is given in section~\ref{sec-out}.

\section{Rotation in rapidly accreting SMSs}
\label{sec-sms}

\begin{figure}\begin{center}
\includegraphics[width=0.7\textwidth]{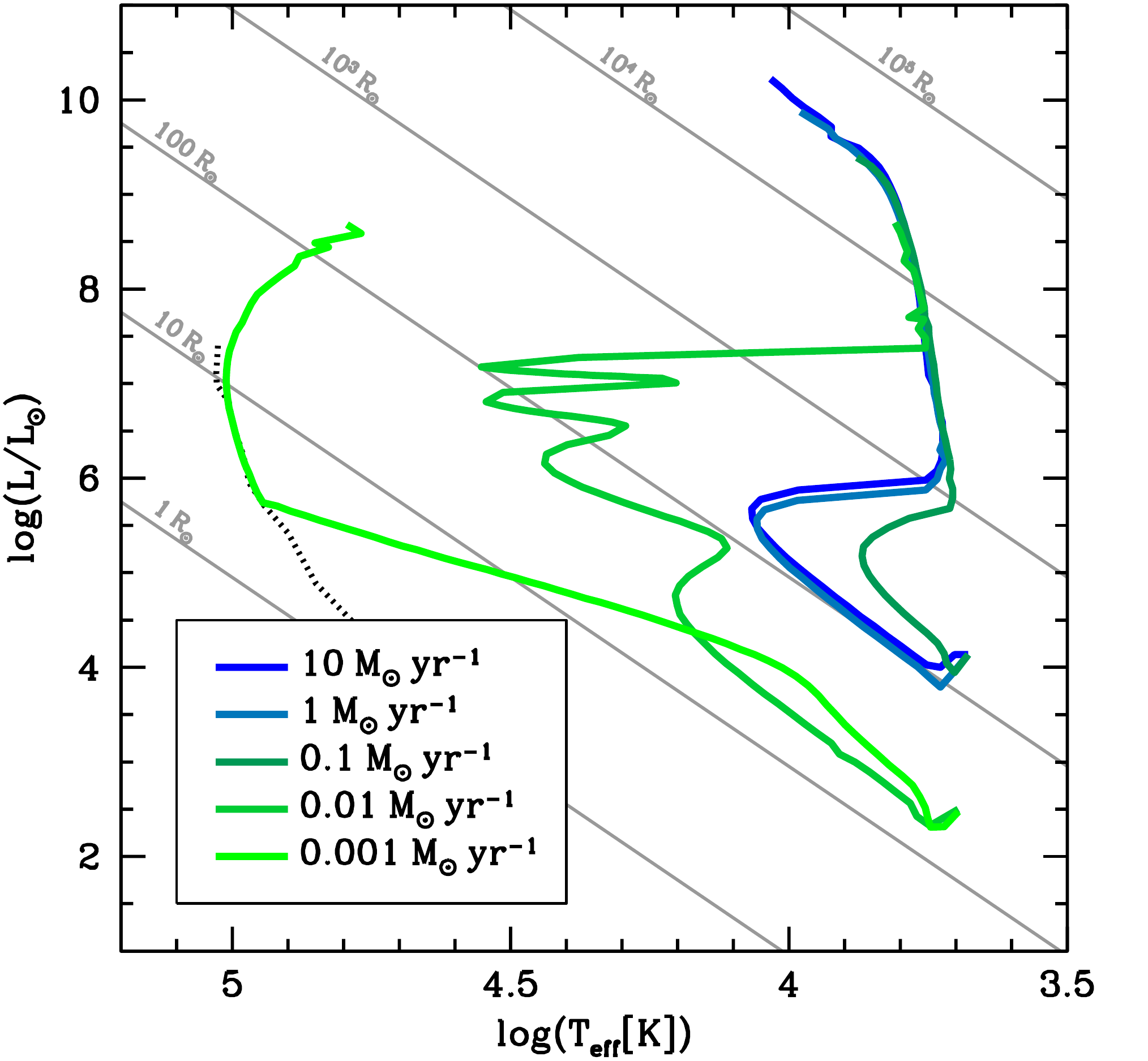}
\caption{Hertzsprung-Russel diagram of the GENEC models of Pop III, rapidly accreting SMSs. (Figure from \cite{haemmerle2018b}.)}
\label{fig-hr}
\end{center}\end{figure}

\begin{figure}\begin{center}
\includegraphics[width=0.6\textwidth]{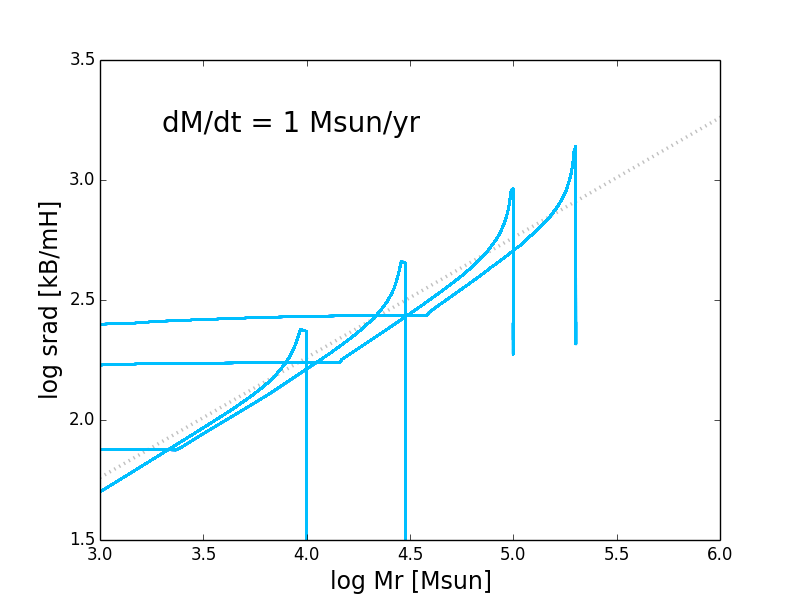}
\includegraphics[width=0.6\textwidth]{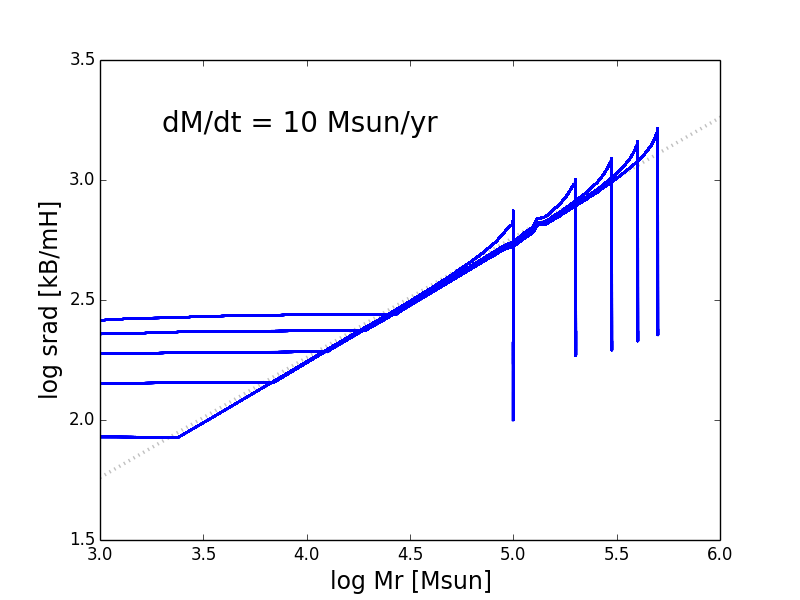}
\includegraphics[width=0.6\textwidth]{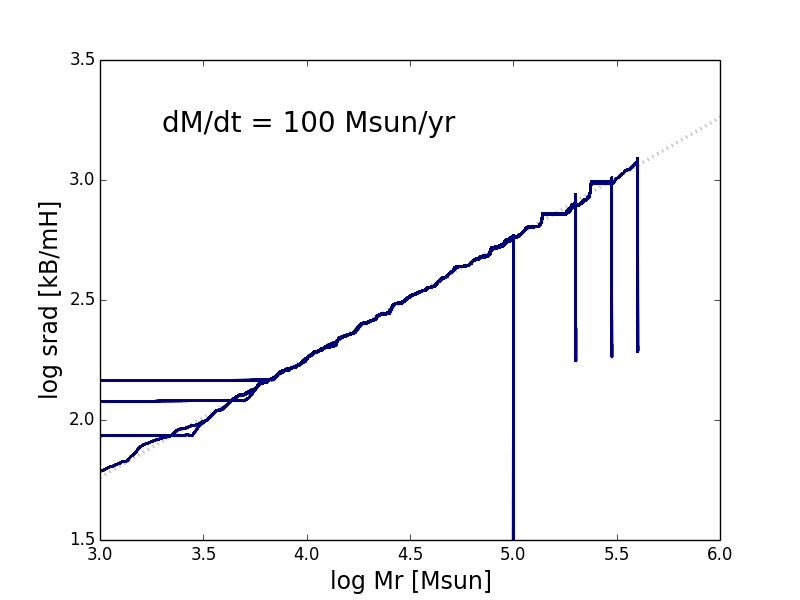}
\caption{Entropy profiles of the GENEC models of Pop III SMSs accreting at $\dot M=1-100$~\Mpy.
The grey dotted line indicates the power-law~(\ref{eq-hylo}).
(Figure from \cite{haemmerle2020b}.)}
\label{fig-hylo}
\end{center}\end{figure}

The evolution of SMSs under rapid accretion has been addressed with analytical and numerical models
in the last decade\cite{begelman2010,hosokawa2013,haemmerle2018b}.
SMSs are always close to the Eddington limit, which implies that their H-burning time is of the order of the Myr, independently of their mass.
Thus, forming a SMS of mass $M\gtrsim10^5$ \Ms\ by accretion before fuel exhaustion requires rates $\dot M\gtrsim0.1$ \Mpy.
Under such accretion rates, SMSs are found to evolve as 'red supergiant protostars'\cite{hosokawa2013},
following upwards the Hayashi limit in the Hertzsprung-Russel diagram.
An example is shown in figure~\ref{fig-hr}, with numerical models of Population III (Pop III) SMSs
computed with the stellar evolution code GENEC\cite{haemmerle2018b}.
The effective temperature on the Hayashi limit remains $\lesssim10^4$ K,
which implies negligible ionising feedback and allows for continuous accretion up to at least $M\sim10^5$ \Ms.
As a consequence of the evolution along both the Eddington and the Hayashi limits,
the mass-radius relation of rapidly accreting SMSs can be approximated by a simple power-law:
\begin{equation}
R=260\,\Rs\left({M\over\Ms}\right)^{1/2}
\label{eq-mr}\end{equation}
The increase of the radius with the growing mass results from the fact that, at rates $\gtrsim0.01$ \Mpy,
the entropy advected by an element of mass $\Delta M$ at accretion cannot be radiated efficiently in the corresponding accretion time $\Delta M/\dot M$.
Thus, the Kelvin-Helmholtz contraction of the stellar interior is too slow compared to the increase in radius that results from the accretion of new mass.
As a consequence, most of the stellar interior is not thermally relaxed and maintains an outwards positive entropy gradient that stabilises the gas against convection.
This is illustrated in figure~\ref{fig-hylo} with the entropy profiles of the Pop III GENEC models at $\dot M=1-100$~\Mpy,
taken at successive stages of the evolution.
The highly centralised energy production by H-burning drives convection in a core of $\sim10\%$ of the stellar mass, that stays isentropic.
An outer envelope is also convective (non-adiabatic), due to the high opacities at the low temperatures of the Hayashi limit.
This outer envelope contains negligible mass ($\sim1\%M$), but represents a significant fraction of the stellar radius.
For accretion rates $\dot M\gtrsim10$~\Mpy, the evolutionary timescale becomes so short that the radiative envelope contracts adiabatically,
compressed by the weight of the layers newly accreted\cite{haemmerle2019c}.
In this regime, the entropy profile converges towards a simple power-law of the mass-coordinates ('hylotrope'\cite{begelman2010}):
\begin{equation}
s\propto M_r^{1/2}
\label{eq-hylo}\end{equation}
This power-law is shown by a grey dotted line in figures~\ref{fig-hylo}.

\begin{figure}\begin{center}
\includegraphics[width=0.6\textwidth]{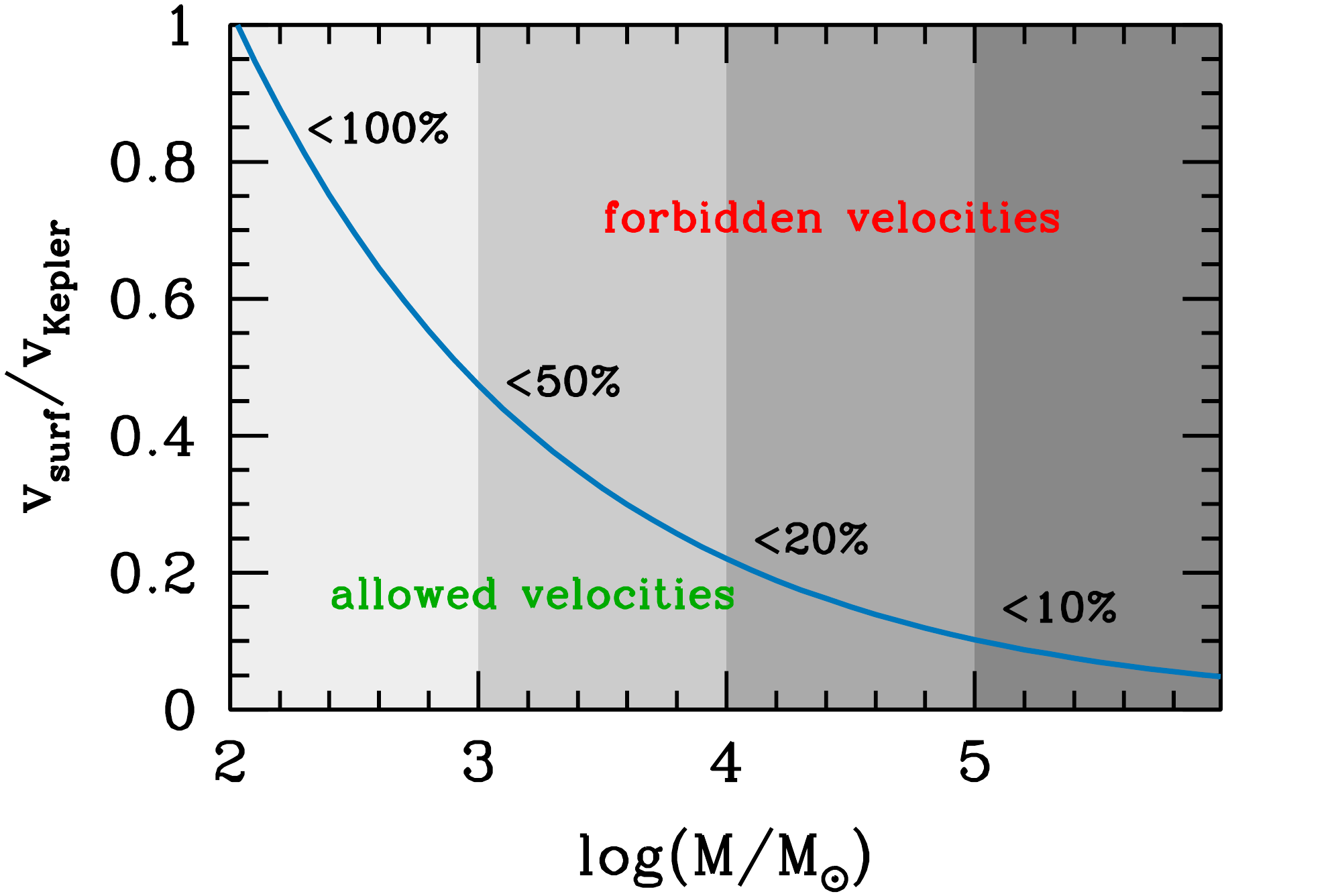}
\caption{Maximum rotation velocity of SMSs as a function of their mass (figure from \cite{haemmerle2018a}).}
\label{fig-omgam}
\end{center}\end{figure}

For stars near the Eddington limit, the maximum rotation velocity consistent with hydrostatic equilibrium is given by the $\Omega\Gamma$-limit,
that accounts for the role of radiation pressure\cite{maeder2000}:
\begin{equation}
\Omega_\Gamma\simeq\Omega_K\sqrt{1-\Gamma}
\end{equation}
where $\Omega_K$ is the Keplerian angular velocity and $\Gamma$ the Eddington factor.
SMSs have $\Gamma\sim0.99$, which implies rotation velocities of $\lesssim10\%\ \Omega_K$;
in other words, SMSs must be slow rotators\cite{haemmerle2018a}.
The maximum rotation velocity set by the $\Omega\Gamma$-limit is shown in figure~\ref{fig-omgam} as a function of the SMS's mass.
The ratio of centrifugal to gravitational forces in the star, which is of the same order as the ratio of rotational to gravitational energies,
can be estimated by
\begin{equation}
{\Omega^2\over\Omega_K^2}\sim1-\Gamma\lesssim1\%
\label{eq-omega2}\end{equation}
For such slow rotations, the deformation of the star by the anisotropic centrifugal force remains negligible.
Moreover, due to the short evolutionary timescales, meridional currents and shear diffusion have no significant impact on the angular momentum distribution.
Thus, the radiative layers contract with local angular momentum conservation, which translates into highly differential rotation,
with a rotation frequency several orders of magnitude larger in the core than at the surface\cite{haemmerle2018a}.
The break-up velocity is not reached in the core, thanks to the large Keplerian velocity in the densest regions.

\section{The angular momentum problem}
\label{sec-J}

\begin{figure}\begin{center}
\includegraphics[width=0.65\textwidth]{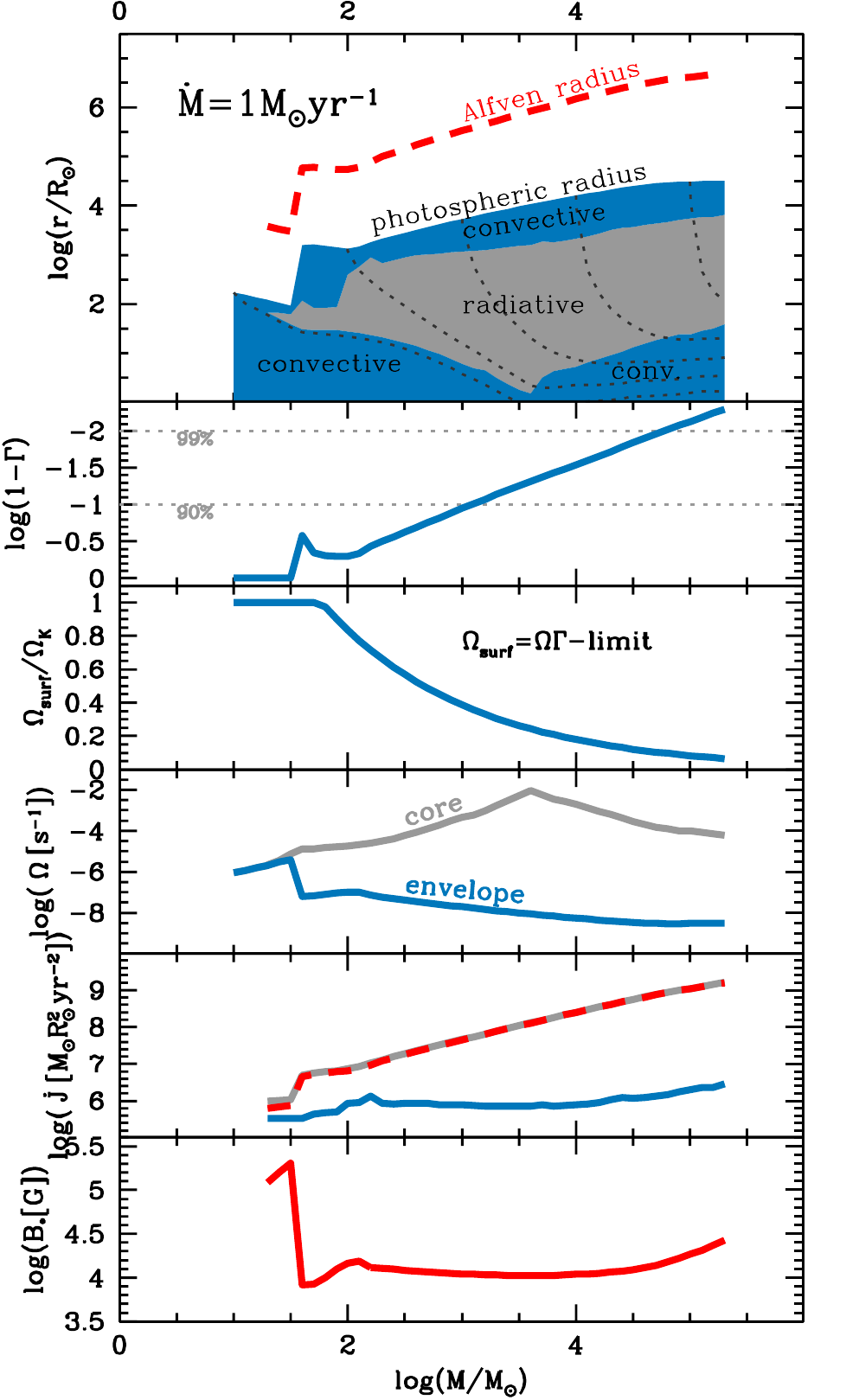}
\caption{GENEC model of a maximally rotating Pop III SMS accreting at $\dot M=1$ \Mpy.
The horizontal axis represents the stellar mass, which is a time coordinate in case of constant accretion.
The first four panels show the stellar structure in radial coordinates, the Eddington factor, the ratio of the surface velocity to the Keplerian velocity,
and the angular velocity in the core and the envelope.
The fifth panel shows the accreted angular momentum (in blue), compared to the Keplerian angular momentun (in grey).
The difference is shown by the red dashed curve.
The bottom panel shows the surface magnetic field required to remove this difference of angular momentum by star-wind coupling.
(Figure from \cite{haemmerle2019a}.)}
\label{fig-mag}
\end{center}\end{figure}

Because of the slow rotations imposed by the $\Omega\Gamma$-limit, SMS formation by accretion implies low angular momentum in the accreted gas.
This is illustrated in figure~\ref{fig-mag},
that shows the evolution of a Pop~III SMS accreting at $\dot M=1$ \Mpy\ under the assumption of maximal rotation.
Rotation is post-processed from the non-rotating GENEC structures shown in the upper panel.
Angular momentum is advected at accretion in order to satisfy the constraint from the $\Omega\Gamma$-limit (third panel) at each time-step,
and the resulting angular momentum accretion history is shown in the fifth panel as a blue line.
It remains 2 -- 3 orders of magnitude below the Keplerian angular momentum (grey line),
which means that, for the $\Omega\Gamma$-limit to be satisfied all along the accretion phase,
the angular momentum advected at accretion cannot exceed a fraction $f\sim0.1-1\%$ of the Keplerian angular momentum.
In other words, SMS formation by accretion from a Keplerian disc requires mechanisms efficient enough
to remove more than 99\% of the angular momentum from the disc.
Notice that the fraction $f$ is 1 -- 2 orders of magnitude smaller than the ratio $\Omega/\Omega_K$ imposed by the $\Omega\Gamma$-limit,
which reflects the instantaneous redistribution of angular momentum in the outer convective envelope visible in the upper panel of figure~\ref{fig-mag}.
Indeed, without convection, the ratio $\Omega/\Omega_K$ in the surface layer
would be simply given by the fraction $f$ of angular momentum advected by this layer.
But here, the surface velocity is given by the angular momentum accreted plus that received by convection from the deeper layers of the envelope.
Thus, a given constraint on the ratio $\Omega/\Omega_K$ (e.g.~$\lesssim10\%$) translates into a tighter constraint on the ratio $f$ (e.g.~$\lesssim0.1-1\%$).

The angular momentum problem is general to star formation\cite{bodenheimer1995}.
For low-mass stars, the problem is easily solved by the magnetic coupling of the star with its disc and winds,
thanks to the convective dynamo in the stellar envelope.
But the problem becomes more difficult for massive stars, since these stars have no convective envelope,
and only a few percent of them have detected magnetic fields, thought to be fossil fields\cite{grunhut2017}.
The most efficient processes to extract angular momentum in the context of massive star formation
rely on the gravitationnal instabilities that develop in massive accretion discs\cite{hosokawa2016}.
Since these gravitationnal instabilities are enhanced by high densities, their efficiency increases when the centrifugal barrier is reached,
as gas accumulates in the vicinity of the star.
Thus, these processes allow for a self-regulated, efficient angular momentum transport all along the collapse.

In the case of rapidly accreting, maximally rotating SMSs,
the surface magnetic field required to remove the angular momentum excess at accretion from a Keplerian disc
by star-winds coupling is of the order of 10 kG, as shown in the bottom panel of figure~\ref{fig-mag}.
The corresponding Alfv\'en radius is shown as a red dashed line in the top panel.
According to estimates based on the Rossby number, a convective dynamo in the envelope of the SMS cannot produce a field larger than 1 G.
Only fossil fields from primordial origin could reach the required values.
On the other hand, gravitational instabilities are particularly strong in the massive accretion discs of SMSs,
and are thought to be efficient enough to circumvent the centrifugal barrier in the context of SMBH formation\cite{begelman2006,wise2008}.

\section{GR instability}
\label{sec-gr}

The formation of a SMBH by direct collapse occurs when the progenitor SMS becomes dynamically unstable.
SMSs are expected to collapse via the GR instability\cite{chandrasekhar1964},
which is a pulsation instability that arises from the GR corrections to the equation of radial momentum.
For a non-rotating post-Newtonian SMS, the eigenfrequency $\omega$ of adiabatic pulsations can be obtained with\cite{haemmerle2021}
\begin{equation}
\omega^2I=\int\beta P\dif V-\int\left({2GM_r\over rc^2}+{8\over3}{P\over\rho c^2}\right){GM_r\over r}\dif M_r
\label{eq-gr}\end{equation}
where $r$ is the radial distance from the stellar centre, $M_r$ the mass-coordinates (i.e. the mass contained in all the layers below $r$),
$\rho$ the mass density, $P$ the pressure, $\beta$ the ratio of gas pressure to total pressure, $I$ the moment of inertia, $V$ the volume of the sphere $r$,
$G$ the gravitational constant and $c$ the speed of light.
Notice that in the post-Newtonian approximation, distinctions between rest-mass and relativistic-mass are irrelevant.

The GR instability is reached when $\omega$ becomes imaginary, which implies that exponentials are solutions to the equations of motion.
Equation~(\ref{eq-gr}) expresses this condition in terms of the two integrals at the right-hand side.
The first one is Newtonian and positive, and scales with $\beta$, that is with the departures from the Eddington limit,
at which all the pressure support is provided by radiation.
As the SMS grows in mass, it approaches the Eddington limit asymptotically with $\beta\to0$, but $\beta>0$.
Thus, without any additional term, equation~(\ref{eq-gr}) would imply that the star remains always stable.
However, the second term, that scales with the GR corrections $2GM_r/rc^2$ and $P/\rho c^2$, is negative.
As the SMS grows in mass, these corrections grow until the sum of the integrals become negative.
At this point, the star is dynamically unstable, which leads in general to the direct formation of a SMBH in a dynamical time\cite{sun2017,sun2018}.

\begin{figure}\begin{center}
\includegraphics[width=0.8\textwidth]{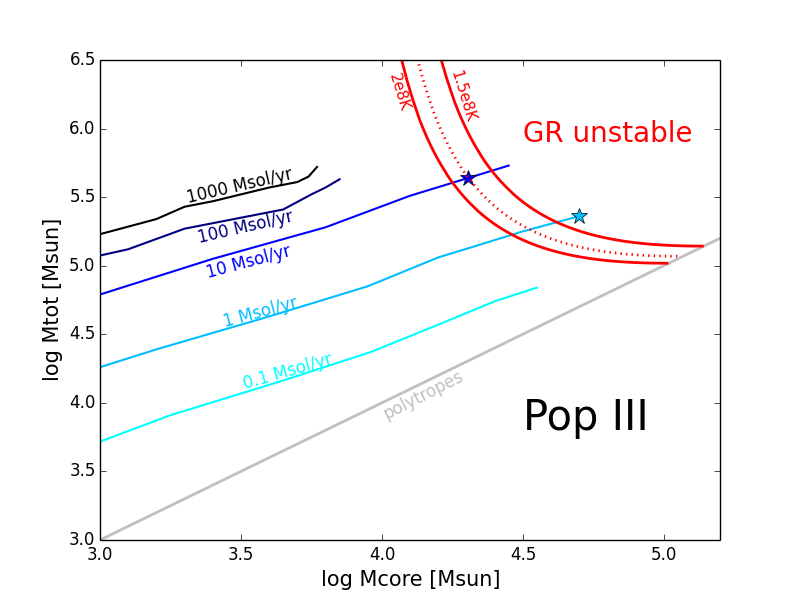}
\caption{Maximum masses of non-rotating Pop~III SMSs consistent with GR stability, as a function of the mass of their convective core.
The red curves show the limit of stability derived from hylotropic structures with the indicated central temperatures.
The black-blue-cyan tracks are Pop~III GENEC models computed with the indicated accretion rates.
The star-like symbols indicate the point of GR instability of these models, when it is reached.
(Figure from \cite{haemmerle2020b}.)}
\label{fig-mmcore}
\end{center}\end{figure}

The exact mass at which the SMS reaches the GR instability depends on its structure, in particular on the mass of its convective core.
The maximum masses of non-rotating Pop~III SMSs as a function of the mass of their convective core are shown in figure~\ref{fig-mmcore}.
The black-blue-cyan tracks are the evolutionary models of Pop~III SMSs computed with GENEC
under accretion at the indicated rates\cite{haemmerle2018b,haemmerle2019c}.
These models were computed under the assumption of hydrostatic equilibrium, which makes them insensitive to dynamical instabilities,
and the ends of the runs are arbitrary, imposed by the numerical stability of the models.
The onset of the GR instability on these hydrostatic models, determined with equation~(\ref{eq-gr}), is indicated by star-like symbols.
It is reached before the end of the run only for rates 1 and 10 \Mpy.
The limitations due to the numerical instability of hydrostatic models with rapid accretion
can be circumvented with the use of semi-analytical models built on the hylotropic law of equation~(\ref{eq-hylo})\cite{begelman2010}.
With these structures, the limit of stability in the $(M_{\rm core},M)$-diagram is uniquely given by the central temperature of the star and its chemical composition.
The central temperature of SMSs is well constrained by numerical models,
and remains in a narrow range due to the thermostatic effect of H-burning
($T_c=1.5-2\times10^8$ K for Pop~III SMSs; $T_c=0.6-0.9\times10^8$ K for Pop~I SMSs).
The limits of stability for Pop~III SMSs derived from the hylotropic models are shown as red curves in figure~\ref{fig-mmcore}.
The dotted red curve is the stability limit for the exact central temperature of the last stable model of the GENEC track at $\dot M=10$ \Mpy.
We see that the hylotropic limit is already very precise for this rate,
in spite of the small departures from the hylotropic profiles visible in figure~\ref{fig-hylo}.
The entropy profile in the inner part, where the density is high and GR corrections are significant,
is already well approximated by the hylotropic law (\ref{eq-hylo}) for this rate.
Only for rates $\lesssim1$ \Mpy\ the GENEC models departs significantly from the hyloptropic law.
But even in this case the discrepancies in the final mass remains of $\sim0.1$ dex, that is $\sim20$ \%.
Thus, the limit of stability given by the hylotropic models remains a good approximation for any accretion rate.
Moreover, the numerical simplicity of these models allows for a broader view of the parameter space.
In particular, we see that for the central temperatures set by H-burning the GR instability cannot be reached if the stellar mass does not exceed $\gtrsim10^5$~\Ms.
Moreover, the GR instability always requires the convective core to contain a mass $\gtrsim10^4$~\Ms.

Rotation plays a prominent role in the stability of SMSs.
Near the GR instability, the dimensionless ratios $\beta$, $2GM_r/rc^2$ and $P/\rho c^2$ are all of the order of a percent.
It implies that the positive and negative integrals at the right-hand side of equation~(\ref{eq-gr})
represent typically a percent of the total internal and gravitational energies of the star, respectively.
Thus, other small effects, that have negligible impact on the hydrostatic structure of SMSs, can play a critical role on the dynamical stability of these structures.
This is the case for rotation: the slow rotation velocities ($\sim0.1\ \Omega_K$) consistent with the $\Omega\Gamma$-limit
correspond to rotational energies of the order of $\Omega^2/\Omega_K^2\sim1\%$ of the gravitational energy of the star (equation~\ref{eq-omega2}).
The rotational corrections to equation~(\ref{eq-gr}) are thus of the same order as the $\beta$- and GR-terms.
For these slow rotations and small GR corrections, however, the relativistic rotation terms are expected to be of second order,
and can be neglected in the determination of the stability limit.
Including the Newtonian rotation term to equation~(\ref{eq-gr}), we obtain\cite{haemmerle2021}
\begin{equation}
\omega^2I=\int\beta P\dif V+{4\over9}\int{\Omega^2\over\Omega_K^2}{GM_r\over r}\dif M_r
-\int\left({2GM_r\over rc^2}+{8\over3}{P\over\rho c^2}\right){GM_r\over r}\dif M_r
\label{eq-gr2}\end{equation}
The positive rotation term contributes to the stability of the star,
and allows it to increase its mass by orders of magnitude beyond the non-rotating limit\cite{fowler1966,haemmerle2021}.

\begin{figure}\begin{center}
\includegraphics[width=0.9\textwidth]{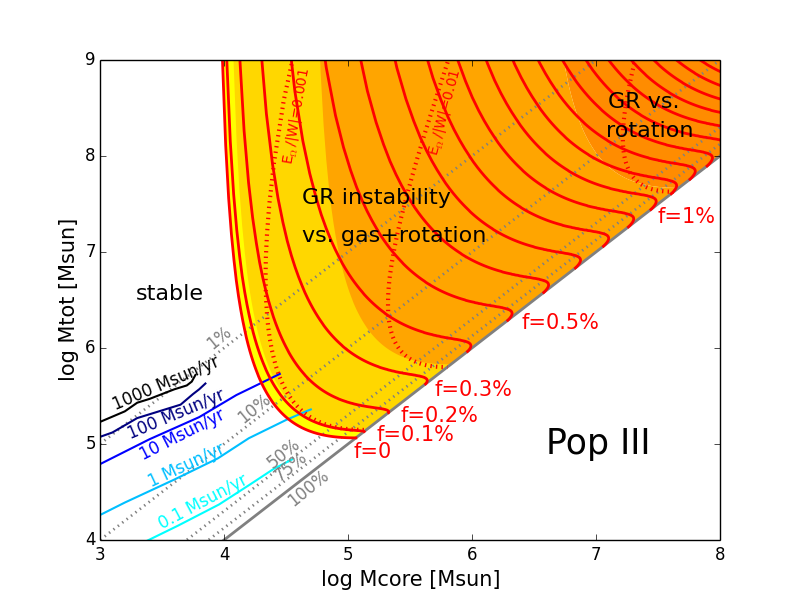}
\caption{Limits of GR stability for rotating Pop~III SMSs, derived from the hylotropic models with $T_c=1.8\times10^8$ K.
The diagram represents the mass of the SMS as a function of the mass of its convective core.
The various limits are obtained with the indicated values of the fraction $f$ of accreted angular momentum over the Keplerian angular momentum (solid red lines).
The limits are also shown for constant ratios of rotational to gravitational energies (dotted red lines).
The coloured areas denote the relative importance of the $\beta$ term (gas term) and the rotation term in the stability of the star (equation~\ref{eq-gr2}).
The black-blue-cyan tracks are the same GENEC models as in figure~\ref{fig-mmcore}.
(Figure from \cite{haemmerle2021}.)}
\label{fig-mmrot}
\end{center}\end{figure}

\begin{figure}\begin{center}
\includegraphics[width=0.9\textwidth]{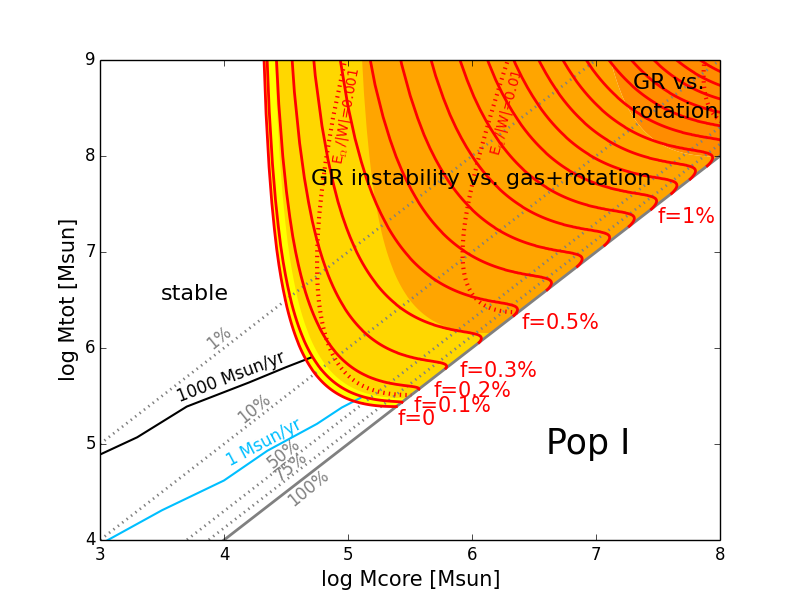}
\caption{Same as figure~\ref{fig-mmrot} for the Pop~I case ($T_c=0.85\times10^8$ K).
(Figure from \cite{haemmerle2021}.)}
\label{fig-mmsol}
\end{center}\end{figure}

The limits of stability for rotating Pop~III and Pop~I SMSs are shown in figures~\ref{fig-mmrot} and \ref{fig-mmsol}, respectively\cite{haemmerle2021}.
The limits have been derived from hylotropic structures,
on which rotation profiles are defined by local angular momentum conservation in the radiative envelope and solid rotation in the convective core.
Each layer of the radiative envelope is assumed to keep the angular momentum it advected at accretion,
defined as a faction $f$ of the Keplerian angular momentum at the accretion radius given by equation~(\ref{eq-mr}).
The angular momentum of the core is then given by that advected from the envelope.
We see that the limits start to be shifted by rotation as soon as $f\gtrsim0.1\%$,
while for $f\sim1\%$ they are already shifted by 3 orders of magnitudes towards larger masses.
The lower central temperatures of the Pop~I models allows for slightly larger masses when $f\sim0.1\%$,
but when $f\sim1\%$ the limits do not depend anymore on the thermal properties of the star, and are uniquely given by $f$.
In this regime, indicated by the red areas on the diagrams, the gas term in equation~(\ref{eq-gr2}) becomes negligible compared to the rotation and GR terms,
which means that rotation becomes the main stabilising agent against the destabilising GR corrections.
An interesting consequence is that the profile of the spin parameter at the onset of the instability, which is key regarding black hole formation,
becomes 'universal', that is, uniquely given by the mass-fraction of the convective core.

For typical conditions of atomically cooled halos (Pop~III, $\dot M\lesssim1$ \Mpy), the $\Omega\Gamma$-limit imposes $f\lesssim0.1-0.2\%$ (figure~\ref{fig-mag}).
In this case, rotation can increase the final stellar mass by up to a factor $\sim2$, which does not allow to exceed $10^6$ \Ms.
As mentioned in section~\ref{sec-J}, this low value of $f$ results from the instantaneous angular momentum transport in the deep convective envelope.
Interestingly, the depth of the envelope decreases significantly for larger rates,
and values $f\sim1\%$ appear consistent with the conditions of merger-driven direct collapse.
In this formation channel, masses up to $10^8-10^9$ \Ms\ could be reached before the GR instability.
Since masses $>10^6$ \Ms\ are required for SMBH formation in the Pop~I case,
the black hole seeds in merger-driven direct collapse must have masses $10^6-10^9$ \Ms,
which implies that the different channels of direct collapse lead to distinct ranges of seed's mass.

\section{SMBH formation and observational signatures of direct collapse}
\label{sec-smbh}

The final collapse of SMSs has been followed with a large number of GR magneto-hydrodynamical simulations\cite{sun2017,sun2018,montero2012}.
In the Pop~III case, the collapse is found to lead to the direct formation of a SMBH, that contains $\gtrsim90\%$ of the total stellar mass.
A fraction $\sim1-10$ \% of the mass can remain in orbit outside the horizon at the formation of the black hole, provided rapid enough rotation.
This gas further collapses through the horizon, with possible gravitational wave emission detectable by future space-based observatories.
Moreover, in the presence of magnetic fields, relativistic jets could be launched and trigger ultra-long gamma-ray bursts.
Thus, the final collapse of SMSs provides the most promising possibilities for observational signatures of direct collapse and the existence of SMSs.

\begin{figure}\begin{center}
\includegraphics[width=0.7\textwidth]{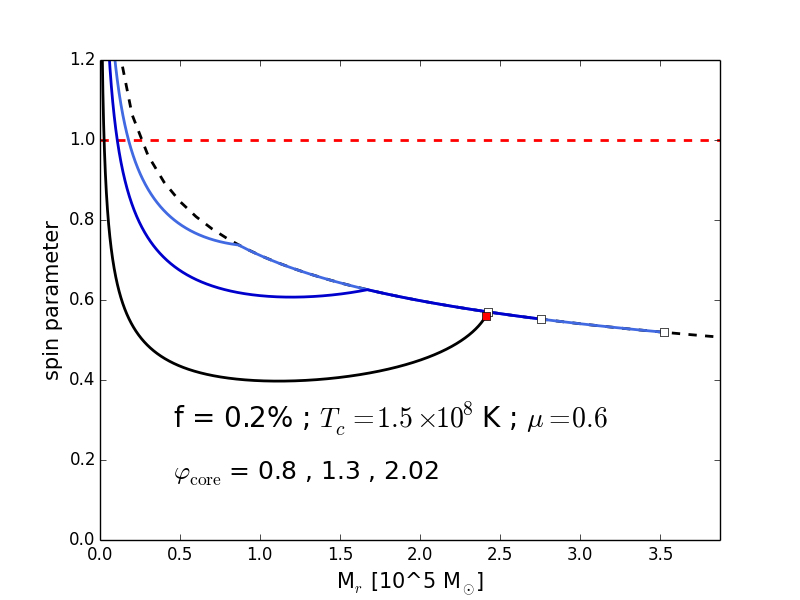}
\caption{Profiles of the spin parameter $cJ_r/GM_r^2$ at the onset of GR instability, for typical conditions of atomically cooled halos.
The profiles (solid lines) are shown for three hylotropes with different core mass-fraction $M_{\rm core}/M$:
the black profile corresponds to a fully convective star $M_{\rm core}=M$ (polytrope),
the dark blue profile corresponds to $M_{\rm core}/M=60\%$ and the light blue profile to $M_{\rm core}/M=25\%$.
The white squares indicate the surface (i.e. the total mass) of the three models (the polytrope has the lowest mass, $M_{\rm core}/M=25\%$ the highest).
Only for the fully convective model a comparison with ISCO angular momentum
suggests that a tiny fraction of the outer mass ($\sim1\%M$) remains in orbit outside the horizon at black hole formation,
which is indicated by the red square.
The black dashed line shows the angular momentum accretion law.
The red dashed line indicates the maximum value of the spin parameter consistent with Kerr black holes.
(Figure from \cite{haemmerle2021}.)}
\label{fig-spin02}
\end{center}\end{figure}

\begin{figure}\begin{center}
\includegraphics[width=0.7\textwidth]{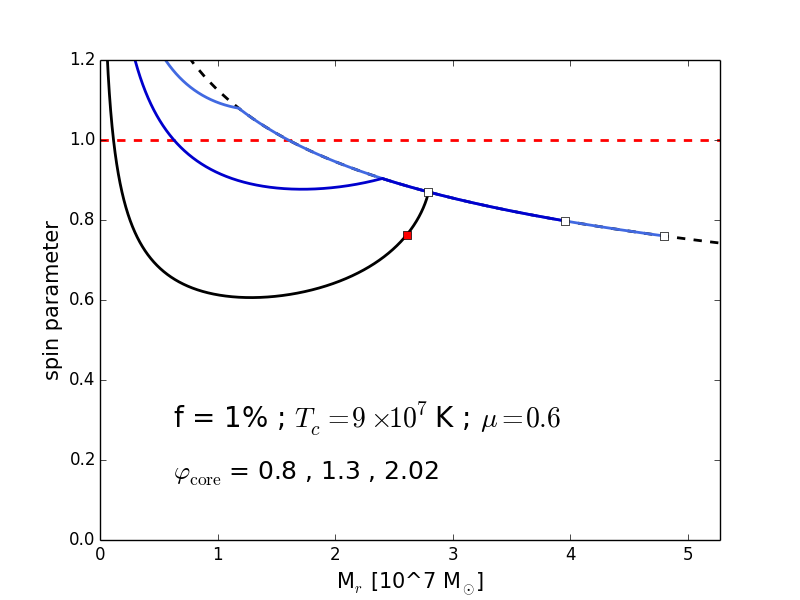}
\caption{Same as figure~\ref{fig-spin02} for typical conditions of merger-driven direct collapse.
Here, the outer mass fraction of the fully convective model that remains in orbit at black hole formation represents $\sim10\%$ of the stellar mass (red square).
(Figure from \cite{haemmerle2021}.)}
\label{fig-spin1}
\end{center}\end{figure}

Rotation is key since only the centrifugal barrier allows for the survival of a slowly decaying torus,
and prevents the direct engulfment of the whole stellar mass by the black hole in a dynamical time.
Figures~\ref{fig-spin02} and \ref{fig-spin1} show the profiles of the spin parameter $cJ_r/GM_r^2$ (where $J_r$ is the angular momentum of the mass $M_r$)
for the hylotropic models of figures~\ref{fig-mmrot} and \ref{fig-mmsol}, taken at the onset of the instability.
They correspond to typical conditions of atomically cooled halos and merger-driven direct collapse, respectively.
In each case, the profiles are shown for three different core mass.
The convective core is recognisable as the only region where the profiles departs from each other, as a result of the angular momentum transport by convection.
In contrast, the profiles match each other in the radiative envelope, reflecting the angular momentum accretion law.
We see that the solid rotation of the core maintains the angular momentum distribution decentralised.
Notice that the profiles of figure~\ref{fig-spin1} correspond to the 'universal' profiles obtained when rotation becomes the dominant stabilising agent.

From these rotation profiles, the fraction of the outer mass that has enough angular momentum to remain in orbit at the formation of the black hole
can be estimated analytically by comparison with the angular momentum of the innermost stable circular orbits (ISCOs)\cite{baumgarte1999}.
It is found to be significant only for the conditions of merger-driven direct collapse, and for fully convective SMSs,
as indicated by the red squares in figures~\ref{fig-spin02} and \ref{fig-spin1}.
The low angular momentum ($f\sim0.2\%$) required in the context of atomically cooled halos prevents the formation of an orbiting structure,
and even for $f\sim1\%$ convection is required in the whole stellar interior for the angular momentum to be sufficiently decentralised.
Overall, these models suggest that gravitational wave emission and ultra-long gamma-ray bursts at black hole formation
are consistent with the conditions of galaxy mergers, but not with those of atomically cooled halos.
In both scenarios, the centrifugal barrier is inefficient to prevent the direct formation of a massive black hole seed at the collapse of the supermassive stellar progenitor.

\section{Summary and conclusions}
\label{sec-out}

SMSs represent promising candidates for the progenitors of the SMBHs that power the most massive quasars detected at redshifts 6 -- 7,
since direct collapse appears as the most efficient scenario to form SMBHs in short timescales.
This scenario faces several bottlenecks, however, but solutions have been proposed to all of them.
The angular momentum problem represents certainly one of the strongest, as discussed in section~\ref{sec-J}.

In spite of the low spins imposed by the $\Omega\Gamma$-limit (section~\ref{sec-sms}, figure~\ref{fig-omgam}), rotation impacts critically the life and death of SMSs.
As an outstanding effect, it allows rapidly accreting SMSs to increase their final masses by several orders of magnitude compared to the non-rotating case,
before they reach the GR instability and collapse into a SMBH (section~\ref{sec-gr}, figures~\ref{fig-mmrot}-\ref{fig-mmsol}).
According to hylotropic models, the masses of the black hole seeds belong to distinct ranges depending on the direct collapse channel:
\begin{itemize}
\item $10^5\ \Ms\lesssim M\lesssim10^6\ \Ms$ for atomically cooled halos;
\item $10^6\ \Ms\lesssim M\lesssim10^9\ \Ms$ for merger-driven direct collapse.
\end{itemize}
Moreover, the conditions of merger-driven direct collapse are expected to be more favourable than those of atomically cooled halos
for gravitational wave emission and ultra-long gamma-ray bursts
at the collapse of the SMS (section~\ref{sec-smbh}, figures~\ref{fig-spin02}-\ref{fig-spin1}).

Many uncertainties remain, however, in these scenarios, most of them relying on the lack of consistency of the models for the formation of the black hole seed itself.
Due to their hypothetical nature, SMSs have been much less studied than other classes of stars,
and the picture given here deserves to be refined by additional models accounting for full stellar evolution up to the largest masses and accretion rates,
including a self-consistent treatment rotation, multiplicity and radiative feedback.
In particular, reliable theoretical predictions for the observational signatures of SMBH formation by direct collapse
require precise determination of the properties of the stellar progenitor.
In that respect, the future generation of space-based telescopes and gravitational wave observatories
could unveil the most massive stars ever formed in Universe's history.
Interestingly, according to the models described here, the maximum mass of these stars is set by the interplay between general relativity and rotation.

\end{document}